\documentclass[
reprint,
superscriptaddress,
 amsmath,amssymb,
 aps,
]{revtex4-2}

\usepackage{graphicx}
\usepackage{dcolumn}
\usepackage{bm}

\usepackage{siunitx}
\usepackage{mhchem}
\usepackage{csquotes}
\usepackage{graphicx}
\usepackage{bm}
\usepackage{mathtools}
\usepackage{braket}
\usepackage{xcolor}

\begin{document}

\preprint{APS/123-QED}

\title{Gate-driven Switching Dynamics in a Fully Suspended Superconducting Nanowire}

\author{Matthijs Rog}
\thanks{These authors contributed equally.}
\affiliation{Huygens-Kamerlingh Onnes Laboratory, Leiden University, 2300 RA Leiden, The Netherlands}

\author{Giacomo Trupiano}
\thanks{These authors contributed equally.}
\affiliation{Istituto Nanoscienze-CNR, NEST-SNS, Piazza San Silvestro 12, 56127 Pisa, Italy}

\author{Filippo Antola}
\affiliation{Istituto Nanoscienze-CNR, NEST-SNS, Piazza San Silvestro 12, 56127 Pisa, Italy}

\author{Mario Cuoco}
\affiliation{CNR-SPIN, Fisciano 84084, Italy}

\author{Giorgio De Simoni}
\affiliation{Istituto Nanoscienze-CNR, NEST-SNS, Piazza San Silvestro 12, 56127 Pisa, Italy}

\author{Kaveh Lahabi}
\affiliation{Huygens-Kamerlingh Onnes Laboratory, Leiden University, 2300 RA Leiden, The Netherlands}

\author{Francesco Giazotto}
\email{francesco.giazotto@sns.it}
\affiliation{Istituto Nanoscienze-CNR, NEST-SNS, Piazza San Silvestro 12, 56127 Pisa, Italy}


\begin{abstract}
A gate voltage can suppress the critical current of a superconducting nanowire, and various scenarios have been proposed to describe these observations. 
The key difficulty is that these scenarios, including both leakage-driven and field-driven mechanisms, are impossible to disentangle in standard experiments.
Here, we resolve this issue by studying a fully suspended superconducting nanowire of amorphous MoGe, separated from the gate by vacuum, eliminating all leakage-driven channels. 
We measure the phase slip escape rate $\Gamma$ and find a large gate-induced enhancement of $\Gamma$, exponential over six orders of magnitude. 
Fitting various escape-rate models reveals a strong deviation from simple phase slip models, hinting at a complex interaction between the electrostatic field and the phase slip centers. 
Surprisingly, $\Gamma$ is not invariant under reversal of the gate-voltage polarity. 
Our work establishes a clear and nontrivial connection between gate voltage and $\Gamma$, guiding future research into the origins of the effect.
\end{abstract}


\maketitle

\section{Introduction}
\label{sec:introduction}
The microscopic mechanism by which a local gate voltage can quench the critical current of a superconducting nanowire remains elusive despite a decade of intensive investigation \cite{de2018metallic, ruf2024gate}. The central question in this debate is whether this gate-controlled superconductivity (GCS) effect is due to the electric field or driven by a small leakage current between the gate and sample. Various leakage-driven effects have been proposed, including Fowler--Nordheim cold-electron field emission from the gate to the wire \cite{golokolenov2021origin}; local heating due to leakage currents \cite{ruf2024high, elalaily2024switching}; and out-of-equilibrium dynamics induced by hot-electron leakage \cite{basset2021gate, elalaily2023signatures}. At the same time, several experiments have pointed to a direct relationship between the electric field and the reduction of the critical current \cite{rocci2020gate, paolucciElectrostaticFieldDrivenSupercurrent2021}. The main obstacle to a comprehensive understanding of the GCS effect is the immense difficulty of untangling the mechanisms at play since multiple channels can, and often do, act \textit{simultaneously}. 

Although understanding the quench of the superconducting state is the ultimate goal of GCS research, $I_\mathrm{c}$-reduction is not the only observed field effect. A gate-driven increase in the phase slip rate was first suggested in experiments on gated constriction junctions \cite{paolucciMagnetotransportExperimentsFully2019}, and later established by a series of experiments that directly probed the phase slip rate \cite{puglia2020electrostatic, basset2021gate, puglia2022phase, elalaily2023signatures}. These studies reveal an anomalous enhancement of the phase-slip rate far below $I_\mathrm{c}$, indicating a reduced critical current. However, almost all of these studies were performed in gated superconducting weak links rather than nanowires, and all were performed on substrates that allow (small) leakage currents. Given the proposed scenarios for leakage-induced phase slips \cite{basset2021gate}, it is therefore crucial to study phase dynamics in leakage-free nanowires under high electrostatic fields.

In this work, we address this need directly by studying the GCS effect in a fully suspended superconducting nanowire of the amorphous superconductor MoGe. Our devices, etched into a silicon-rich nitride (SiRN) membrane, feature a \SI{70}{\nano\metre} vacuum gap between gate and nanowire that rules out substrate-mediated leakage, as demonstrated previously in an experiment on a suspended Ti nanowire \cite{rocci2020gate}. In this system, we find that applying the gate does not simply rescale the critical current. Instead, application of the gate broadens the switching-current distribution by inducing phase slips at currents far below $I_\mathrm{c}$. Several striking properties of these phase-slip-rate curves show that they cannot be trivially connected to phase slips at zero gate. First, we observe an exponential dependence of the phase slip rate on the applied gate voltage. Second, we observe an anomalous phase-slip rate exponent irreconcilable with standard theories of uncorrelated phase slips. 
Third, the rate is asymmetric in the gate-voltage polarity, although it is fully symmetric in the bias current. From this we conclude that the gate acts directly on the escape dynamics of the superconductor rather than as a static and uniform perturbation of the condensate.

\begin{figure*}[!t]
\includegraphics{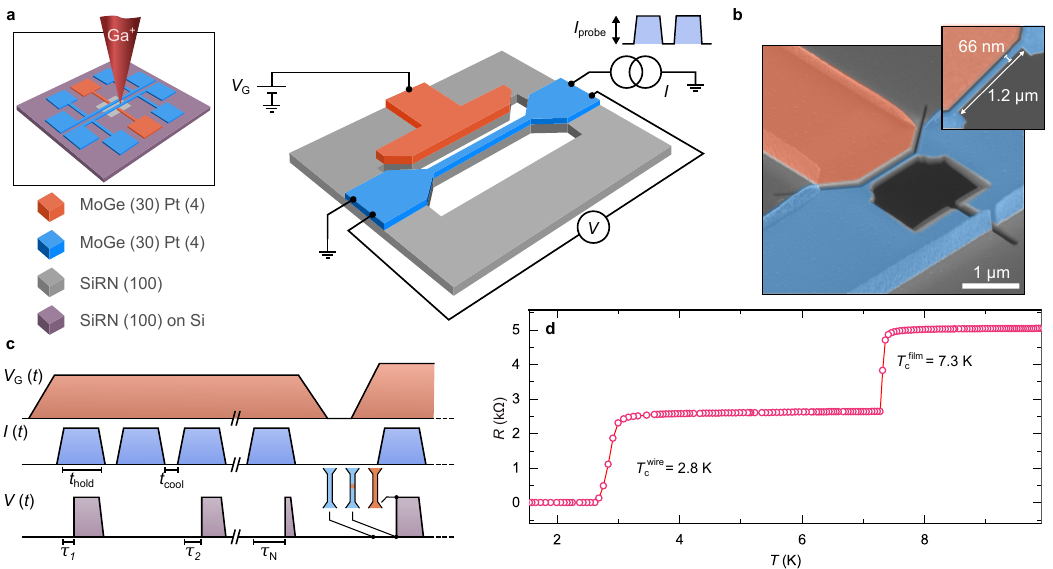}
\caption{\label{fig:Fig1}(a) Illustration of the experimental geometry. Using a $\mathrm{Ga}^\mathrm{+}$ focused ion beam (inset), a MoGe nanowire is structured on top of a SiRN membrane. 
Milling through the membrane and the MoGe separates the nanowire (blue) from the gate (orange) by a \SI{70}{\nano\metre} vacuum gap, eliminating any substrate- or dielectric-mediated leakage path. (b) False-colored scanning electron micrograph of the nanowire used in this work. The inset shows the wire's dimensions. (c) Experimental protocol for measuring the gate-induced phase slip rate $\Gamma$. First, the gate voltage $V_\mathrm{G}$ is slowly ramped to its setpoint. Then, the nanowire current $I$ is very quickly ramped to its setpoint $N$ times. The switching times $\tau_1, \tau_2, ..., \tau_\mathrm{N}$ are then recorded. (d) Resistance-temperature curve of the nanowire in (b) showing two transitions. The wire becomes superconducting at $T_\mathrm{c}^\mathrm{wire}= \SI{2.8}{\kelvin}$.}
\end{figure*}

Unlike crystalline or polycrystalline metals studied in the gate-effect literature to date, amorphous MoGe has no long-range crystalline order. Because this categorically excludes contributions of intergrain Josephson coupling \cite{arutyunovSuperconductivityOneDimension2008}, MoGe has emerged as the material of choice for controlled studies of quantum and thermal phase slips in superconducting nanowires \cite{sahu2009individual, aref2012quantitative}.

In a current-biased nanowire, switching to the resistive state is a stochastic event, set by the rate $\Gamma$ at which the wire escapes its metastable state through a phase slip, analogous to the escape dynamics first characterized for current-biased Josephson junctions \cite{fulton1974lifetime}. Phase slips can be thermally activated over a current-dependent free-energy barrier \cite{langer1967intrinsic, mccumber1970time}, or occur via macroscopic quantum tunneling \cite{giordano1988evidence} at low temperatures. Consequently, the switching current is determined stochastically through this escape rate rather than by a sharp threshold. In GCS, the phase-slip rate $\Gamma$ has so far been inferred indirectly, from switching-current probability distributions accumulated over repeated bias ramps \cite{puglia2020electrostatic, puglia2022phase}, by way of the Kurkij\"arvi-Fulton-Dunkleberger inversion \cite{kurkijarvi1972intrinsic, fulton1974lifetime}.

Initially motivated by the broadening of the switching-current distribution observed in the DC $IV$ characteristics, we instead measure $\Gamma$ directly from fixed-bias waiting-time statistics. This approach avoids the numerical differentiation inherent to the distribution-based method and is particularly well suited to resolving the low-probability tails central to the results below.

\section{Results and Discussion}
\label{sec:results-discussion}

\begin{figure*}[t]
\includegraphics{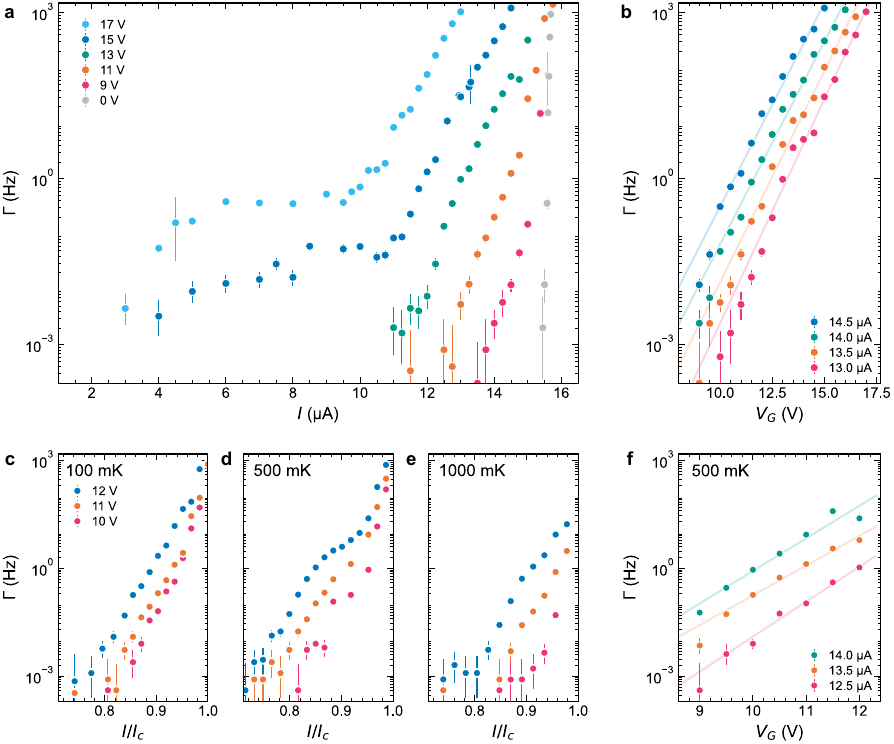}
\caption{\label{fig:Fig2} (a) Gate-induced phase slip rates $\Gamma$ as function of bias current at \SI{100}{\milli\kelvin} of bath temperature. Already at $V_\mathrm{G}\sim\SI{9}{\volt}$, the phase slip curve broadens to lower currents. For high gate voltages $V_\mathrm{G} \geq$ 15 V, a pronounced tail develops that stretches all the way to $\sim 0.2 I_\mathrm{c}$. (b) Constant-current phase-slip rate curves as a function of gate voltage at a bath temperature of \SI{100}{\milli\kelvin}. The solid lines show exponential fits $\Gamma \propto \exp(a V_\mathrm{G})$ to the data, demonstrating that the rate-voltage relation is exponential to high accuracy over six orders of magnitude. (c-e) Phase slip rate curves at three different temperatures. The effect is present at all temperatures over an approximately equal range of currents (relative to $I_\mathrm{c}$), although the detailed shape of the curve changes with temperature. (f) At \SI{500}{\milli\kelvin}, the same exponential dependence between $\Gamma$ and $V_\mathrm{G}$ appears. To good accuracy, the (logarithmic) slope of this relation is the same for all currents, and corroborates that determined in (b).}
\end{figure*}

Figure~\ref{fig:Fig1} shows the device and the measurement protocol used to investigate the gate-induced escape dynamics. As shown in Fig.~\ref{fig:Fig1}a-b, a \ce{MoGe} nanowire and an adjacent side gate are etched by $\mathrm{Ga}^\mathrm{+}$ focused-ion-beam milling directly through a suspended SiRN membrane, so that both the wire (blue) and the gate (orange) are released from the substrate and fully separated by a vacuum gap. This intrinsically excludes the substrate-mediated channels invoked in the literature to explain the GCS effect: diffusive leakage currents, phonon heating, and quasiparticle injection are geometrically ruled out. The resistance-temperature curve of the wire in Fig.~\ref{fig:Fig1}d shows two resistive transitions: one is the transition of the \ce{MoGe} film at $T_\mathrm{c}^\mathrm{film}=\SI{7.3}{\kelvin}$, the second is the transition of the nanowire $T_\mathrm{c}^\mathrm{wire}=\SI{2.8}{\kelvin}$. 

Our measurement protocol is described in Fig.~\ref{fig:Fig1}c. To determine the escape rate $\Gamma$, we first ramp the gate voltage ($V_\mathrm{G}$) slowly to its set point, then ramp the bias current ($I$) quickly to a fixed set point. We maintain this current bias for a set time while monitoring the voltage across the wire. A trial in which the wire remains superconducting for the full period is recorded as a non-escape observation. Otherwise, the time $\tau$ at which the wire switches to the resistive state, measured from the end of the current ramp, is recorded. This ramp-hold-reset cycle is repeated $N$ times (typically several hundred repetitions) at each $(I, V_\mathrm{G}, T)$ setpoint, accumulating a set of switching times $\tau_1,\tau_2,...,\tau_\mathrm{N}$. Between repetitions, we reset the bias current and hold it at zero for a cooling interval much longer than the wire's thermal recovery time. 
We discard trials that switch during the current ramp. To prevent long-term charge buildup, we also periodically cycle the gate to zero. All observations are consistent with Poissonian statistics; therefore, we infer $\Gamma$ from the pooled statistics of all $N$ trials using maximum likelihood estimation \cite{Lawless2003}.



\begin{figure*}[!t]
\includegraphics{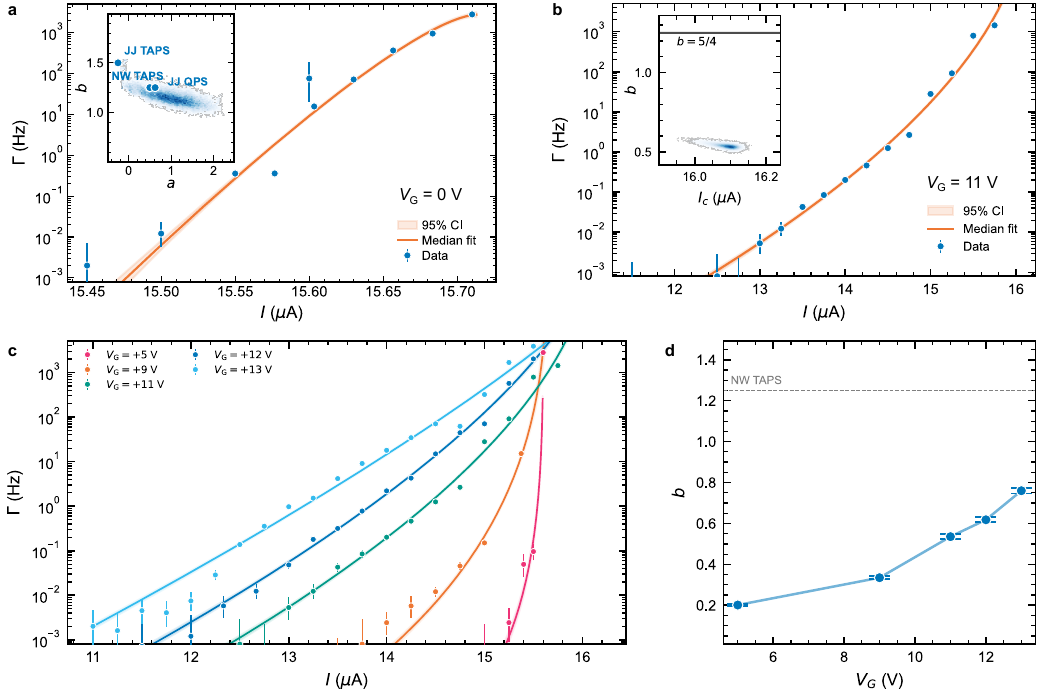}
\caption{\label{fig:Fig3} (a) Markov Chain Monte Carlo (MCMC) fit of the general LAMH form $\Gamma = A (1-I/I_\mathrm{c})^{a+b-1}\exp{[-B(1-I/I_\mathrm{c})^b]}$ \cite{gargEscapefieldDistributionEscape1995} to the $V_\mathrm{G}=0$ escape-rate curve. The inset shows the posterior distributions for the exponents $a$ and $b$. We also label the exponents predicted for thermally activated (TAPS, escape over the barrier) and macroscopic-quantum-tunneling-activated (QPS, escape through it) phase slips in a Josephson junction (JJ) and a nanowire (NW) \cite{gargEscapefieldDistributionEscape1995, murphy2013universal}. The exponents depend only on the underlying escape mechanism and are independent of, for instance, the energy-barrier height. Although the exponent $a$ is constrained very poorly, the fitted exponent $b$ is constrained strongly and agrees with the literature value for phase slips in a nanowire $b=5/4$. (b-c) MCMC fits of Kramers' model to curves of non-zero $V_\mathrm{G}$, with (b) showing the $V_\mathrm{G}=\SI{11}{\volt}$ case and (c) showing the fit results for all curves. For the fits, we fix $a=5/8$ and assume $A$ and $B$ are gate-independent. The model fits the data to high accuracy, but for each escape-rate curve, it finds a \textit{convex} ($b<1$) curve. The inset to (b) displays the posterior distribution of $b$ and $I_\mathrm{c}$. Clearly, both are tightly constrained, and $b<1$ with high certainty. (d) Evolution of the exponent $b$ with $V_\mathrm{G}$. Although the exponent climbs with gate voltage, it never reaches the expected $5/4$ for nanowires.} 
\end{figure*}

Figure~\ref{fig:Fig2} shows the central results of our study. The first panel shows $\Gamma$ as a function of the bias current at fixed temperature (\SI{100}{\milli\kelvin}) for a series of gate voltages, including a zero-gate control measurement. Interestingly, we observe an enhanced phase-slip rate at currents far below $I_\mathrm{c}$. For sufficient gate voltage $V_\mathrm{G}\simeq\SI{17}{\volt}$, this enhancement is present even at $\simeq 20\%$ of $I_\mathrm{c}$ (measured at $V_\mathrm{G}=0$), displaying the remarkable interaction between the application of an electrostatic field and the generation of phase slips. We stress that the resulting $\Gamma(I)$ curves are not simply broadened or rescaled versions of the zero-gate curve. We find below that, in fact, their curvature has the opposite sign to that expected from an increase in effective temperature, leaving $b=5/4$ unchanged.

At fixed bias current, $\Gamma$ increases monotonically with $V_\mathrm{G}$, as shown in Fig.~\ref{fig:Fig2}b. 
Remarkably, this $\Gamma-V_\mathrm{G}$ relation is well described by a single exponential $\Gamma \propto \exp(a V_\mathrm{G})$, and the fits hold to high accuracy over \textit{six} orders of magnitude in $\Gamma$. The lack of systematic deviation from this single-exponential form suggests that the gate acts smoothly on the escape process, rather than through threshold behavior or a crossover between distinct regimes. Such a dependence would be consistent with the gate acting on the phase-slip free-energy barrier itself, 
and echoes earlier reports of electrostatic control of phase-slip rates in gated all-metallic Josephson nanotransistors~\cite{puglia2020electrostatic,puglia2022phase}. 


The exponential form itself already constrains the mechanism. Considering an Arrhenius rate $\Gamma=\Omega\exp(-\Delta F/k_\mathrm{B}T)$~\cite{langer1967intrinsic, mccumber1970time}, with $\Omega$ an attempt frequency and $\Delta F$ the phase-slip barrier, we can use simple scaling predictions to rule out several scenarios. For instance, the electrostatic force ($\propto V_\mathrm{G}^{2}$) can lower the barrier through the resulting strain, giving $\Gamma\propto\exp(cV_\mathrm{G}^{2})$. Dissipation ($P\propto V_\mathrm{G}^{2}$) instead increases the wire temperature $T_\mathrm{e}$ in the denominator. For the strong heating needed to reach the measured rates and a linear thermal conductance, $T_\mathrm{e}\propto V_\mathrm{G}^{2}$ and $\Gamma\propto\exp(-C/V_\mathrm{G}^{2})$, while cooling channels that grow faster with temperature, such as electron-phonon coupling~\cite{giazotto2006opportunities}, make the dependence even weaker. Here, $c$ and $C$ are positive constants. If each emitted electron triggers switching with a fixed probability, Fowler--Nordheim emission~\cite{rocci2020gate, golokolenov2021origin} gives $\Gamma\propto V_\mathrm{G}^{2}\exp(-V_\mathrm{FN}/V_\mathrm{G})$, with a characteristic voltage $V_\mathrm{FN}$ inversely proportional to the field-enhancement factor $\beta$ (see Appendix~B). The first form makes $\ln(\Gamma)$ convex in $V_\mathrm{G}$ and the other two concave, so over the six orders of magnitude probed each would leave clear deviations from the observed exponential behavior. The measured curves show no such deviations, remaining linear within the scatter across the full range. For Fowler--Nordheim emission, this conclusion holds for any $\beta$.

In addition, a direct estimate of the Fowler--Nordheim current quantitatively rules out cold-electron field emission. The field-enhancement factor $\beta$, set by the relative length scale of deviations and protrusions from the surface, is the key parameter. Sputtered amorphous films naturally have an extremely low roughness. 
We then take, as a worst-case estimate, the generous value $\beta \approx 5$ from standard field-enhancement models \cite{forbes2003enhancement} (see Appendix~B). A Fowler--Nordheim~\cite{golokolenov2021origin, rocci2020gate} estimate using this $\beta$ for our geometry predicts a field-emission current of order $10^{-48}$~A at $V_\mathrm{G}=9\,\mathrm{V}$, rising to $10^{-26}$~A at $V_\mathrm{G}=17\,\mathrm{V}$. This shows that even at the highest gate voltages probed, the probability of a single emission event over the acquisition time of the \textit{complete dataset} is vanishingly small.


We repeat these measurements at three temperatures, shown in Fig. \ref{fig:Fig2}c-e. At every temperature studied, the gate-induced enhancement of $\Gamma$ is present over an approximately equal range of bias current relative to $I_\mathrm{c}$. The precise shape of the curve $\Gamma(I)$ changes with temperature. This is expected since the underlying escape process itself is thermally activated at higher temperatures and crosses over to macroscopic quantum tunneling at low temperatures \cite{mccumber1970time, giordano1988evidence, sahu2009individual, aref2012quantitative}, independently of any gate effect. The same exponential gate dependence with a comparable logarithmic slope is recovered at 500 mK, as shown in Fig.~\ref{fig:Fig2}f. 
This indicates that the smooth, single-exponential form of $\Gamma(V_\mathrm{G})$ established above is preserved throughout the temperatures probed, even as the shape of $\Gamma(I)$ itself changes. 

The results so far establish that the gate voltage strongly enhances the phase slip rate. We now turn to constraining the physical mechanism that could mediate this enhancement. To do so, we investigate the precise shape of the $\Gamma(I)$ curves. 

In Langer-Ambegaokar-McCumber-Halperin (LAMH) theory, the escape rate takes the form $\Gamma(I,T) = \Omega(I,T)\exp\left[-\Delta F(I)/k_BT\right]$ \cite{langer1967intrinsic, mccumber1970time}, where $\Omega(I,T)$ is an attempt frequency and the current-biased free-energy barrier takes the form $\Delta F(I) = \Delta F(0)\left[1 - I/I_\mathrm{c}\right]^{b}$. The exponent $b$ depends on the microscopic physics at play: for thermally activated phase slips in a superconducting nanowire, theory predicts $b=5/4$ \cite{Khlebnikov2016}, and experiments have shown this exponent to carry over into the quantum regime \cite{sahu2009individual, aref2012quantitative}. Indeed, fitting the LAMH form to our $V_\mathrm{G} = 0$ data is fully consistent with this prediction, as we show in Fig.~\ref{fig:Fig3}a. Further details of the Monte Carlo fitting procedure are available in Appendix D. 


We subsequently investigate how the gate voltage modifies this exponent. We summarize our fits in Fig.~\ref{fig:Fig3}d. We observe that at $V_\mathrm{G}=5$\,V the exponent drops to $b\approx 0.2$ and remains below 0.8 for all gate voltages probed. These exponents, which produce \textit{convex} curves on a log-linear plot, stand in stark contrast to earlier observations of gate-enhanced phase slips, which yielded concave ($b>1$) curves \cite{basset2021gate}. This suggests that different physical mechanisms operate now that the leakage currents are removed. 

This naturally raises the question of how to interpret $b<1$. The LAMH model was originally developed to describe nanowires. Nevertheless, the same functional form, with similar exponents, also describes Josephson junctions (Fig.~\ref{fig:Fig3}a, inset). In both the thermal and quantum regimes, the theory predicts $b>1$ for nanowires and junctions alike \cite{Khlebnikov2016, fulton1974lifetime, gargEscapefieldDistributionEscape1995}. In other words, for all known model systems, theory predicts a \textit{concave} curve in Fig.~\ref{fig:Fig3}b-c, rather than the observed \textit{convex} behavior. We stress that $b$ as a parameter is completely independent of both $\Delta F(0)$---the zero-current energy barrier---and $I_\mathrm{c}$. Therefore, neither a simple lowering of the effective energy barrier nor a heating-induced decrease in $I_\mathrm{c}$ could explain these observations.

\begin{figure}[t]
\includegraphics{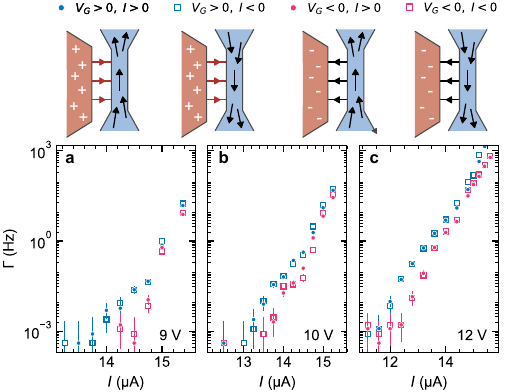}
\caption{\label{fig:Fig4} (a-c) Observation of gate-polarity symmetry breaking for $|V_\mathrm{G}|=9$ V (a), 10 V (b), and 12 V (c). To high accuracy, the phase slip rate is unchanged when the current sign (circles/squares) is inverted. However, when the gate-voltage sign is reversed (blue/magenta), the phase slip rate differs markedly, sometimes by more than an order of magnitude. 
Therefore, the response is not invariant under gate-polarity reversal.
}
\end{figure}

The evolution of the switching-current exponent from the conventional quantum phase-slip value $b=5/4$ at zero gate voltage to $b\approx1/4$ at intermediate gate voltage and subsequently to $b\approx3/4$ at larger gate voltage (Fig.~\ref{fig:Fig3}d) suggests that the switching dynamics cannot be described solely in terms of independent phase-slip events. Motivated by theoretical studies of interacting and correlated quantum phase slips in superconducting nanowires \cite{MooijSchon1985,Zaikin1997,Buchler2004,arutyunovSuperconductivityOneDimension2008,SemenovZaikin2017}, we consider an effective nonlocal action in which the applied gate modifies the electromagnetic environment and thus the interaction between quantum phase slip events. In this framework, the gate controls the correlation length of collective tunneling configurations, renormalizing the switching exponent $b$. In particular, correlated quantum phase slip configurations can reduce the switching exponent to values as small as $b\approx1/4$, while stronger screening at larger gate voltages shortens the effective interaction range and drives the exponent toward $b\approx3/4$. We present the derivation of the non-local action and the associated scaling arguments in Appendix E. We emphasize that this interpretation should be regarded as a phenomenological scenario illustrating how gate-induced quantum phase slip correlations can account for the observed behavior, while alternative mechanisms, including coupling to additional collective modes or other many-body effects, cannot be excluded at the present stage.

\bigskip

Finally, we test the symmetry of the effect by reversing the polarity of the bias current and the gate voltage independently, at three gate-voltage magnitudes: $|V_\mathrm{G}|=$ $9$, $10$ and \SI{12}{\volt}. The results are shown in Fig.~\ref{fig:Fig4}. This experiment reveals a peculiar asymmetry. Reversing only the current direction leaves $\Gamma$ unchanged to high accuracy. In contrast, reversing the sign of $V_\mathrm{G}$ changes $\Gamma$ measurably, sometimes by more than an order of magnitude. 
This directly shows that the response is not invariant under gate-polarity reversal.

This asymmetry is notable against the backdrop of the wider gate-effect literature: critical-current suppression in all-metallic devices has been repeatedly reported to be ambipolar and symmetric under $V_\mathrm{G} \to -V_\mathrm{G}$~\cite{de2018metallic, rocci2020gate, ritter2022out}. This sign-independence is considered a benchmark for candidate mechanisms~\cite{ritter2022out}. The clear symmetry-breaking in this experiment is therefore either a feature that coarser critical-current measurements in earlier work could not resolve, or a genuine distinction introduced by the fully suspended, vacuum-only geometry. A polarity-independent perturbation such as ohmic leakage heating would instead leave $\Gamma$ even in $V_\mathrm{G}$, since the dissipated power driving it scales as $V_\mathrm{G}^2$ and is therefore strictly even. The only polarity-dependent, leakage-driven GCS mechanism is cold-electron field emission across the vacuum gap. This mechanism is clearly asymmetric in $V_\mathrm{G}$: the sign of the gate voltage determines the path of the emitted electrons. Either the electrons move from the gate to the nanowire or from the nanowire to the gate. The former should heat the nanowire considerably more than the latter, since a ballistically accelerated electron deposits its kinetic energy $eV_\mathrm{G}$ where it lands, not where it started. Notice that, in our data, the rate is larger for $V_\mathrm{G}>0$, when electrons would be emitted from the wire into the gate. This is the opposite of the asymmetry expected from field-emission heating. Moreover, considering given the vacuum separation between the gate and the sample, it is not clear that any gate heating should affect the nanowire at all. Even setting this aside, we reiterate that from our quantitative estimate above, field emission across the gap is far too weak to account for the observed escape rate.

A possible microscopic origin of the observed polarity dependence is a local electromagnetic or polarization mode coupled to both the gate electric field and a quantum phase slip event. Integrating over this mode generates a contribution to the effective quantum phase slip action linear in $V_\mathrm{G}$. For two degenerate polarization configurations related by gate reversal, selection of the lower-energy branch converts this contribution into a dependence on $|V_\mathrm{G}|$. A weak asymmetry can arise from a small energy splitting between the two configurations caused by trapped charges or surface disorder. Although not unique, a charged or polarizable surface defect is a plausible microscopic mediator of this coupling between the gate and the quantum phase slip degrees of freedom. We note that coupling to a polarizable collective mode would also naturally give rise to nonlocal interactions between quantum phase-slip degrees of freedom, thereby providing a consistent microscopic mechanism for the behavior of the anomalous switching rate discussed above.

These observations in this geometry favour direct field-driven effects \cite{mercaldo2020electrically,bours2020unveiling,chirolli2021impact,solinas2021sauter}. Even in the realm of field-driven effects, the observation of gate-polarity-symmetry breaking is a crucial benchmark against which theoretical proposals must be gauged. For instance, because of the \enquote{hangbridge} geometry of the device and the high electrostatic fields employed, electromechanical forces are non-negligible. In principle, a gate-voltage dependent strain or deformation could modify the free-energy landscape experienced by the phase slips. However, as elaborated further in Appendix C, the electromechanical forces are strictly even in $V_\mathrm{G}$, ruling out an exclusively electromechanical theory as a full explanation of the observed phenomenology. Further calculations in the same Appendix indicate that the gate-induced change in the gate-wire gap is limited to below $1 \%$, while the largest displacements along the nanowire are approximately \SI{0.6}{\nano\metre}. 


The fact that the effect persists in an amorphous material without long-range crystalline order is notable in itself. In at least one proposed field-driven scenario, crystalline order is specifically invoked to produce the GCS effect \cite{chirolli2021impact}. In sharp contrast, our results show no hard requirements on crystalline order. Although this does not exclude a role for the local or short-range orbital structure, it disfavors mechanisms that rely on the extended crystalline band structure. 

\section{Conclusion}
\label{sec:conclusion}

In conclusion, we have identified a gate-driven enhancement of the phase slip rate in a fully suspended amorphous MoGe superconducting nanowire gated across a vacuum gap. By construction, this device excludes all substrate-mediated leakage effects. Moreover, at the probed field scales, cold-electron field emission remains extremely unlikely. Consequently, we identify a direct, electric field-driven scenario as the leading candidate for the observed phenomena.

The microscopic mechanism through which the electric field affects the phase slip rate remains unclear. However, three striking observations constrain the mechanism. First, the escape rate $\Gamma$ increases by six orders of magnitude as a single exponential in $V_\mathrm{G}$. Second, the $\Gamma(I)$ curves have a log-convex curvature, opposite to that predicted by simple heating, and tentatively indicating a connection to field-induced phase-slip correlations. Lastly, $\Gamma$ is asymmetric in $V_\mathrm{G}$ but not in the bias current $I$, 
in contrast to the approximately ambipolar response reported in earlier literature.

We hope that these observations, as well as the introduction of direct escape-rate 
measurements in suspended geometries, will guide future research into the microscopic origins of the GCS effect. In particular, we foresee that systematic studies of various suspended superconductors will further clarify the extent to which the observed anomalous phase slips are a universal feature of field-driven GCS effects. 

\section*{Acknowledgements}
\label{sec:Acknowledgements}

The authors acknowledge partial financial support from the PNRR MUR project PE0000023-NQSTI. The project was also supported by EU Cost Action CA16218 (NANOCOHYBRI).
We acknowledge Tycho Blom, Reinier Regter, Luc Wigbout, and Sander van Leeuwen for their technical support. 

\bibliographystyle{achemso}
\bibliography{bibliography}

\appendix

\section{Methods}
\begin{figure*}[!t]
\includegraphics{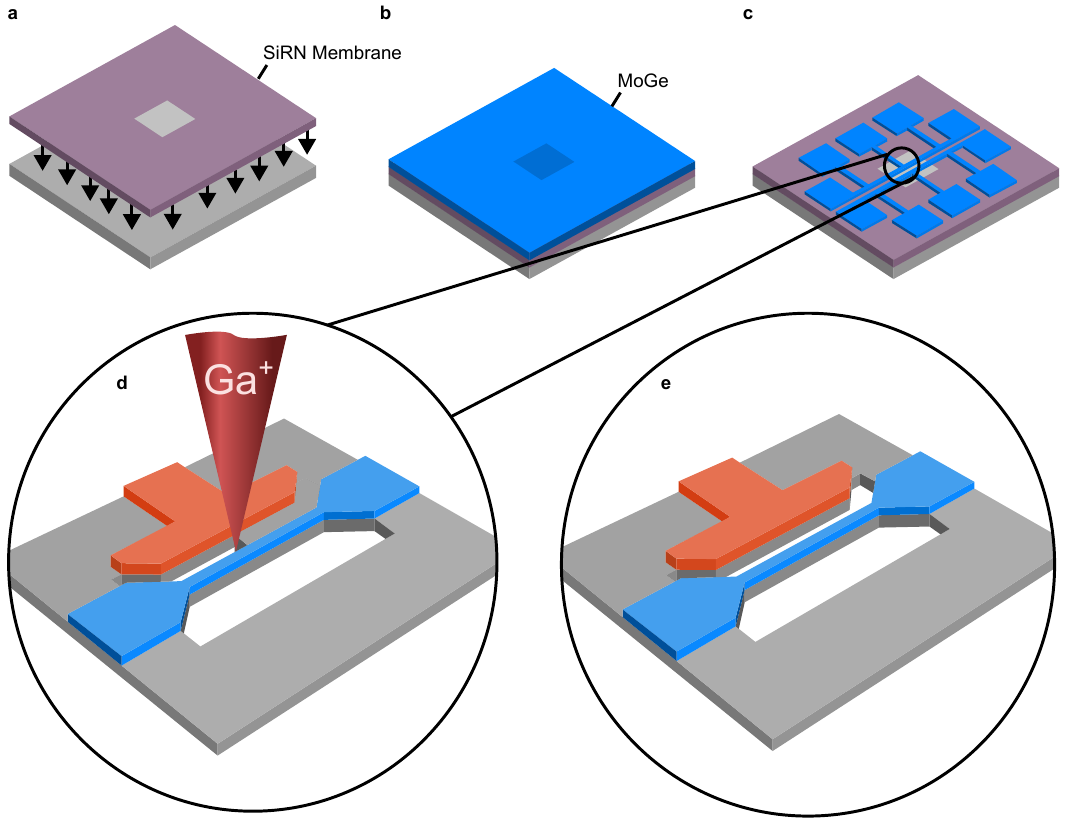}
\caption{\label{fig:FabIllustration}Outline of the procedure to create a suspended MoGe nanowire. (a) The SiRN membrane is glued onto a carrier chip. (b) A MoGe film is sputtered onto the membrane. (c) Electron-beam lithography and Ar-ion milling define structures on top of the membrane. (d) \ce{Ga+} focused ion beam milling produces a fully suspended nanowire. (e) The final sample.}
\end{figure*}
Our fabrication methods are summarized in Fig.~\ref{fig:FabIllustration}. To produce our suspended nanowires, we use \SI{100}{\nano\metre} thick SiRN membranes in a \SI{200}{\micro\metre} silicon frame. We use high-resistivity silicon to prevent spurious leakage currents through the substrate. Before fabrication, we glue the membranes onto a silicon carrier chip to ease manual handling and enable spin-coating. Subsequently, a uniform film (thickness \SI{30}{\nano\metre}) of amorphous MoGe is deposited inside a Leybold Z-400 diode sputtering system. Although MoGe is quite air-stable, we cap all films \textit{in situ} with \SI{4}{\nano\metre} of Pt to protect against degradation and etching. We then use standard electron-beam lithography to define two devices per membrane in CSAR 62 etch-resistant resist. We etch away unnecessary parts of the film using Ar-ion milling. Care is taken to over-etch approximately \SI{25}{\nano\metre} into the SiRN; such an over-etch ensures no spurious MoGe remains on the substrate. We then transfer the samples to a ThermoFischer Aquilos dual-beam electron microscope equipped with a module for \ce{Ga+} focused ion beam milling. Using a series of milling steps, we first separate the gate from the nanowire, then progressively thin the nanowire to the desired width. This procedure gradually releases stress in the membrane, preventing wire buckling.

We measure the devices inside an Oxford Instruments Triton 200 dilution refrigerator with a base temperature of 20 mK. We perform all transport measurements in a standard 4-wire configuration, amplifying the voltage across the wire $100\times$ with a low-noise preamplifier (NF DL1201) before readout. We first characterize transport using a Yokogawa GS200 voltage source across a series $200\,\mathrm{k}\Omega$ resistor as a current source, with an HP 34410A digital multimeter reading the amplified voltage. Next, we perform escape-rate spectroscopy using a NI USB-6281 data acquisition system, which pulses the bias current across the same $200\,\mathrm{k}\Omega$ resistor and digitizes the same amplified voltage signal to record the switching events. The voltage across the wire is digitized at $10$ kSa\,s$^{-1}$. The combination of sampling rate and the rise-time of the employed cryogenic RC-filters enables us to resolve escape rates up to $\sim\SI{1}{\kilo\hertz}$.

\section{Fowler--Nordheim and leakage currents}

Cold-electron field emission follows the Fowler--Nordheim law
\begin{equation} I_\mathrm{FN}=A_\mathrm{e}\,\frac{a}{\phi}\,F^{2}\exp\!\left(-\frac{b\,\phi^{3/2}}{F}\right),\qquad F=\frac{\beta V_\mathrm{G}}{d}, \end{equation}
where $a=e^{3}/(8\pi h)\approx1.54\times10^{-6}$\,A\,eV\,V$^{-2}$ and $b=8\pi\sqrt{2m_\mathrm{e}}/(3eh)\approx6.83$\,eV$^{-3/2}$\,V\,nm$^{-1}$ are the Fowler--Nordheim constants~\cite{Forbes2007}, $\phi$ is the work function and $A_\mathrm{e}$ the area of the emitting surface, $F$ the local field at that surface, $d$ the gate-wire distance and $\beta$ the field-enhancement factor. Hence, $I_\mathrm{FN}\propto V_\mathrm{G}^{2}\exp(-V_\mathrm{FN}/V_\mathrm{G})$ with $V_\mathrm{FN}=b\,\phi^{3/2}d/\beta$. We take the whole $34\,\mathrm{nm}\times1.2\,\mu\mathrm{m}$ sidewall facing the gate as $A_\mathrm{e}$, $d=70$\,nm and the work function of Mo, $\phi=4.6$\,eV~\cite{Michaelson1977}, so that $V_\mathrm{FN}\approx4.7\,\mathrm{kV}/\beta$. 
We take $\beta \approx 5$ as a generous value. The electrostatic problem is scale invariant, so $\beta$ depends on the shape of a protrusion and not on its size~\cite{forbes2003enhancement}: an isolated hemispherical bump on a flat surface gives $\beta = 3$ at any radius, and $\beta \approx 5$ requires a protrusion about three times taller than its radius. 
Supplying $10^{2}$ electrons per second at $V_\mathrm{G}=9$\,V requires $\beta\approx15$, much larger than the assumed $\beta\approx5$. This needs a height-to-radius ratio of about 13, so the larger the radius, the taller the protrusion: over 25 nm for radii above 2 nm, which would be clearly visible across the 70 nm gap in Fig.~\ref{fig:Fig1}b. Only atomically sharp spikes would escape SEM, and these are excluded by the functional-form of the $\Gamma-V_G$ curve argument.

We separately measure a current of about 15\,pA at $V_\mathrm{G}=9$\,V and 30\,pA at $V_\mathrm{G}=17$\,V between the gate and wire lines. Its linear dependence on $V_\mathrm{G}$ ($\approx0.6$\,T$\Omega$) excludes field emission across the gap, which would grow by many orders of magnitude over this range. 
Two lines of the same setup left open at the sample end draw a current of the same order, and currents of this size are routinely measured in our setup. Therefore, we attribute the gate current mainly to the wiring. Even if part of it flowed through the chip, it could not account for the observed effect, since the shape of $\Gamma(I)$ excludes heating and a reduced $I_\mathrm{c}$ (Sec.~II). Discrete leakage-driven events are likewise excluded: their rate would follow the ohmic gate current, which grows by $\sim10\%$ between 15 and 17 V, whereas $\Gamma$ on the current-independent plateaus of Fig. 2a grows by more than an order of magnitude.

\section{Simulation of electromechanical strain}
\begin{figure*}[!t]
\includegraphics{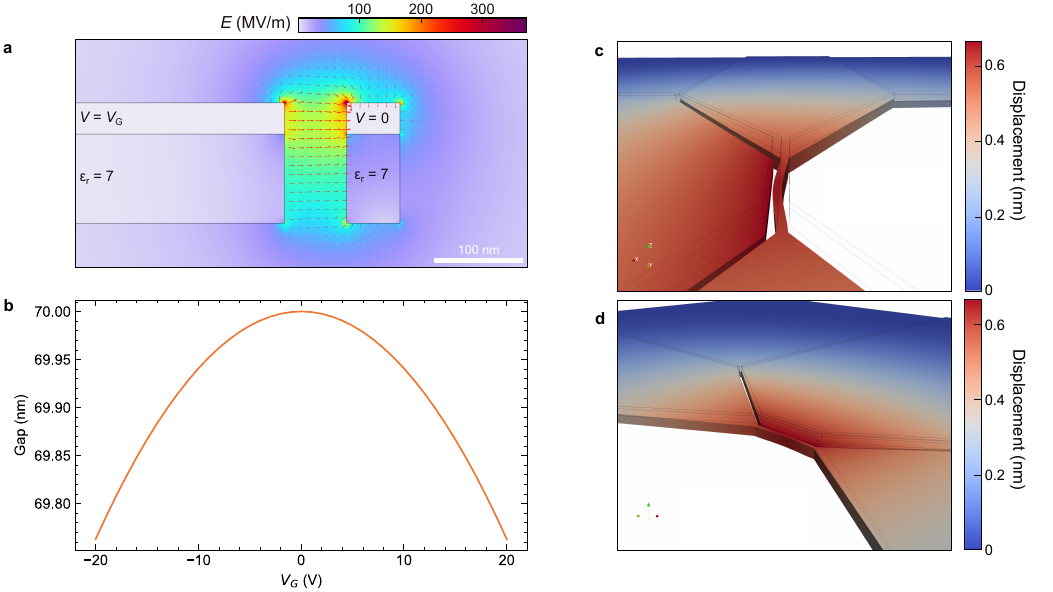}
\caption{\label{fig:ComsolResults}Electromechanical simulation of gate-induced strain and the subsequent displacement of the nanowire. (a) Cross-cut of the geometry, showing the electric field around the gate and nanowire. (b) From an electromechanical model, we extract the precise gap distance between the nanowire and the gate. The resultant displacement is even in gate voltage and never larger than \SI{150}{\pico\metre} for the experiments of the main text. (c-d) Displacement of the nanowire, exaggerated by a factor $300$ for visualization purposes, from two different perspectives.}
\end{figure*}
To rule out strain effects induced by the electric field on the nanowire, we perform a fully three-dimensional simulation in COMSOL Multiphysics. We model a representative nanowire with a gate-wire gap of \SI{70}{\nano\metre} and assume that in equilibrium the gate is perfectly coplanar with the membrane (which minimizes the gate-wire gap). 

We perform the simulations in two steps. First, we compute the electric field around the geometry. We describe the membrane as a \SI{100}{\nano\metre} perfect dielectric with relative permittivity $\epsilon_r = 7$. We model the metal as a perfect conductor of thickness \SI{35}{\nano\metre}, meaning we fix the electric potential on the metal surfaces and do not compute the electric field inside---it is zero. This gives the electric-field distribution in Fig.~\ref{fig:ComsolResults}a. 

Next, we compute the mesh deformation induced by the electric fields on the surfaces. For this, we compute the Maxwell stress tensor in the presence of zero magnetic field

\begin{equation}
    T_{ij} = \epsilon_0 (E_i E_j - \frac12 E^2\delta_{ij})
\end{equation}

The Maxwell stress tensor is even with respect to field polarity. From the $T_{ij}$ tensor, the force per unit volume can be calculated:

\begin{equation}
    \mathbf{f} = \nabla \cdot \mathbf{T}
\end{equation}

Using the divergence theorem, we calculate the force on any area element. We then use a linear elastic model to calculate the displacement of each surface from the known and fixed stress tensor. Because the displacement is much smaller than the characteristic mesh size, we do not need to iterate the electrostatic and elastic models.

Figure~\ref{fig:ComsolResults}c,d show the (scaled-up) displacement of the nanowire in the electric field. The nanowire is attracted towards the gate, while the whole system also bends downward. This downward bending can be traced to the downward force exerted at the \enquote{triple point} where vacuum, dielectric, and metal meet. Figure~\ref{fig:ComsolResults}b shows the extracted change in the gate-wire gap. Compared to equilibrium, all relative changes in the gap are well below 1\%. Thus, these minute deformations are unlikely to cause the observed effect. Furthermore, we re-emphasize that these electrostatic forces are fully ambipolar, showing that the observed asymmetries cannot be described by an electromechanical model. 

\section{Monte Carlo Fitting Procedure}
Here we describe how we obtain the LAMH exponent $b$ from a Markov Chain Monte Carlo (MCMC) fitting procedure.

Following Garg \cite{gargEscapefieldDistributionEscape1995}, we use the general Kramers form of the escape rate $\Gamma$ from a metastable state:

\begin{equation}\label{eqn:kramers}
    \Gamma = A \left(1 - \frac{I}{I_\mathrm{c}}\right)^{a+b-1}
    \exp\left[-B\left(1-\frac{I}{I_\mathrm{c}}\right)^b\right]
\end{equation}

This equation has five unknowns: $A$, whose inverse sets the natural timescale of the model; the exponents $a$ and $b$, which encode the physical mechanism for escape; $B$, which encodes the barrier height; and $I_\mathrm{c}$, the critical current. 

Although this expression captures a wide range of escape-rate phenomena, the large number of parameters creates degeneracy among the fit parameters. For instance, unless many orders of $(1-I/I_\mathrm{c})$ are probed, $A$ and $a$ are degenerate. In principle, one could treat all five parameters as gate-dependent and obtain perfect fits---we have indeed confirmed this separately. However, because of these degeneracies, no physical insights can be extracted from this! In a situation like this---a large number of fitting parameters that are possibly degenerate and connected through non-Gaussian statistics---MCMC fitting strategies are the natural choice. 

We use all curves between $V_\mathrm{G}=\SI{5}{\volt}$ and $V_\mathrm{G}=\SI{13}{\volt}$. For higher gate voltages (cf. Fig. \ref{fig:Fig2}a) the rate curves contain anomalous steps that are impossible to describe with equation \ref{eqn:kramers}. We explicitly do \textit{not} fit all curves independently, to circumvent the problem of degeneracy. Instead, we fix and postulate that a single value of $A$ and $B$ describes all curves---essentially, we hypothesize that applying the gate voltage changes the \textit{physical mechanism for escape} rather than the barrier height. Furthermore, because it is strongly degenerate with $A$ and has only a weak effect on the rate, we fix $a$ at its theoretical value $5/8$ \cite{murphy2013universal}. Only $b$ and $I_\mathrm{c}$ are then curve-dependent. With five independent curves, this gives $12$ free parameters to describe the entire dataset. 

To find these parameters $\mathbf{\theta}=(A,B,b_1,b_2,...)$, we define a standard log-likelihood function:

\begin{equation}
    \mathcal{L} = -\frac{1}{2} \sum_i \left[ \left(\frac{\log \Gamma_i - \log m_i(\boldsymbol{\theta})}{\sigma_i^{\pm}}\right)^2 + \log (\sigma_i^{\pm})^2 \right]
\end{equation}

Here $\Gamma_i$ is the $i$th measured escape rate, and $m_i(\boldsymbol{\theta})$ is the corresponding model prediction. We fit in log-space to weight all decades of $\Gamma$ equally. The asymmetric error bars $\sigma_i^\pm$ on $\log\Gamma_i$ are those obtained in the Poissonian analysis; we select $\sigma_i^+$ or $\sigma_i^-$ depending on the sign of the residual to account for the asymmetry of the error distribution. We additionally impose a Jeffreys prior $p(B)\propto 1/B$ on the barrier height parameter, and uniform priors on all other parameters.

We then optimize the log-likelihood using the standard MCMC procedure, yielding posterior distributions for all $12$ fit parameters. Some (marginal) posterior distributions are shown in the insets to Fig. \ref{fig:Fig3}. 

\section{Switching Current Exponent and Quantum Phase Slips with Gating}

\textit{Gate dependent Switching-Rate Scaling Near the Critical Current.} We consider the evolution of the supercurrent switching rate in
nanowires under an applied gate voltage close to the critical current. We write
\begin{equation}
\Gamma(I,V_\mathrm{G})=\Gamma_0(I,V_\mathrm{G})
\exp\!\left[
-A(V_\mathrm{G})\left(1-\frac{I}{I_c}\right)^{\mu(V_\mathrm{G})}
\right],
\end{equation}
where $\mu(V_\mathrm{G})$ is an effective exponent associated with the quantum
phase-slip (QPS) dynamics. Experimentally, the exponent evolves
approximately as
\begin{equation}
\mu(V_\mathrm{G}):\qquad
\frac54\ \longrightarrow\ \frac14\ \longrightarrow\ \frac34 .
\end{equation}
This non-monotonic evolution suggests that the gate modifies not only
the QPS amplitude but also the interaction between QPS events.

We propose that the gate first promotes a correlated QPS configuration
with a large correlation length and, at larger voltage, progressively
screens the interaction, reducing its spatial range. Interacting and
correlated QPS configurations have been considered in finite
superconducting wires and in theories of QPS-pair formation
\cite{Buchler2004,Belkin2015,arutyunovSuperconductivityOneDimension2008}.

For weak QPS--QPS interaction, the switching dynamics can be described
by a single collective QPS coordinate $\phi$ in a tilted washboard
potential,
\begin{equation}
L=\frac{M}{2}\dot{\phi}^{\,2}-U(\phi),\qquad
U(\phi)=-g\cos\phi-gs\phi,\qquad s=\frac{I}{I_c}.
\end{equation}
At low temperature the switching rate is determined by the Euclidean
bounce,
\begin{equation}
\Gamma_Q=A_Qe^{-S_B/\hbar},
\quad
S_B=2\int_{\phi_m}^{\phi_t}d\phi\,
\sqrt{2M[U(\phi)-U(\phi_m)]},
\end{equation}
with $\phi_m$ the metastable minimum of the potential and $\phi_t$ the classical turning point.
Close to the critical current, defining
$\epsilon=1-I/I_c\ll1$, the barrier and characteristic tunnelling time
scale as
\begin{equation}
\Delta U\propto\epsilon^{3/2},
\qquad
\tau_B\propto\epsilon^{-1/4},
\end{equation}
giving
\begin{equation}
S_B\sim\Delta U\,\tau_B\propto\epsilon^{5/4}.
\end{equation}
Thus a single-QPS mechanism naturally produces
\begin{equation}
\Gamma_Q\sim
A_Q\exp\!\left[-S_0
\left(1-\frac{I}{I_c}\right)^{5/4}\right],
\end{equation}
consistent with the low-gate-voltage regime
\cite{GolubevZaikin2001,Khlebnikov2016}.

At finite gate voltage, QPS events can interact through the
electromagnetic and charge modes of the nanowire
\cite{MooijSchon1985,MooijNazarov2006,arutyunovSuperconductivityOneDimension2008,SemenovZaikin2017}.
Integrating out these modes generates an effective nonlocal QPS
interaction,
\begin{equation}
V_{\rm corr}=
\frac{\kappa}{2}\int d\tau\,dx\,dx'\,
K(x-x';V_\mathrm{G})
[\phi(x,\tau)-\phi(x',\tau)]^2 ,
\end{equation}
where the kernel $K$ is determined by the electromagnetic environment.
A nearby gate modifies the electrostatic boundary conditions and can
therefore screen the interaction, providing a mechanism for tuning its
effective spatial range.

The corresponding action is
\begin{equation}
S_E=\int d\tau dx
\left[
\frac{M}{2}(\partial_\tau\phi)^2+U(\phi;s)
\right]+V_{\rm corr},
\end{equation}
with saddle-point equation
\begin{equation}
-M\partial_\tau^2\phi+
g(\sin\phi-s)
+2\kappa\!\int\!dx'\,K(x-x')
[\phi(x)-\phi(x')]=0 .
\end{equation}
Rather than solving this nonlocal instanton problem numerically, we use
a scaling form
\begin{equation}
K(x)\propto |x|^{-1-\sigma}.
\end{equation}
Near $I_c$, the instanton amplitude and temporal scale obey
\begin{equation}
\phi_{\rm char}\sim\epsilon^{1/2},
\qquad
\tau_B\sim\epsilon^{-1/4},
\end{equation}
while balancing the nonlocal interaction with the driving term gives
\begin{equation}
L_B\sim\epsilon^{-1/(2\sigma)}.
\end{equation}
Consequently,
\begin{equation}
S_B\sim
\epsilon^{3/2}\epsilon^{-1/(2\sigma)}
\epsilon^{-1/4}
=
\epsilon^{5/4-1/(2\sigma)}.
\end{equation}
For $\sigma=1/2$ this gives
\begin{equation}
S_B\propto\epsilon^{1/4},
\qquad
\Gamma_Q\propto
\exp[-S_0\epsilon^{1/4}],
\end{equation}
providing a possible description of the intermediate-gate regime.
Increasing $\sigma$ corresponds to a more rapidly decaying, effectively
shorter-ranged interaction and can increase the effective exponent,
potentially producing values approaching $3/4$ at larger gate voltage.
Thus the observed
$5/4\rightarrow1/4\rightarrow3/4$ evolution can be interpreted
phenomenologically as a gate-controlled crossover between weakly
interacting, strongly correlated, and more strongly screened QPS
configurations.

\textit{Gate-linear contribution to the QPS action.}
To account for the experimentally observed linear-in-$V_\mathrm{G}$ contribution to the exponential dependence of the switching rate, one must identify a microscopic mechanism that produces a term proportional to $V_\mathrm{G}$ in the QPS action. A natural possibility is a local polarization or electromagnetic mode $X$ that couples simultaneously to the QPS configuration and to the electric field generated by the gate. 
A minimal Hamiltonian can be the following
\begin{equation}
H=
\frac{\Pi_\phi^2}{2M}+U(\phi;I)
+\frac{P_X^2}{2m_X}
+\frac{m_X\omega_X^2X^2}{2}
-\lambda XF(\phi)-p\beta V_\mathrm{G}X ,
\end{equation}
with the electric field $E=\beta V_\mathrm{G}$ and $p$ the polarization of the mode.
Eliminating $X$ gives
\begin{equation}
H_{\rm eff}=
\frac{\Pi_\phi^2}{2M}+U(\phi;I)
-\frac{\lambda^2F^2}{2m_X\omega_X^2}
-\frac{\lambda p\beta}{m_X\omega_X^2}V_\mathrm{G}F(\phi)
+\mathrm{const.}
\end{equation}
and hence a term in the QPS action proportional to $V_\mathrm{G}$. Such a mode
could represent screening charge near the QPS core, a charge
redistribution mode, or a polarizable surface defect.

If the polarization mode has two nearly degenerate configurations
related by reversal of the electric field, minimization over the two
states produces an approximately $|V_\mathrm{G}|$ dependence and therefore a
bipolar switching response. A small built-in field, trapped charge, or
surface asymmetry would lift the degeneracy and naturally produce a
weak polarity asymmetry. This provides a possible microscopic mechanism
for the observed gate dependence of the QPS action.


\end{document}